\documentclass[twocolumn,floats,floatfix,superscriptaddress,aps,pra]{revtex4}
\usepackage{subfigure}
\usepackage{psfrag,graphicx}
\usepackage{bm,color}
\usepackage{latexsym}
\usepackage{epstopdf}
\usepackage[english]{babel}
\usepackage{amsmath,amssymb,amsfonts}
\usepackage{appendix}
\usepackage{bbold}
\usepackage[dvipsnames]{xcolor}

\definecolor{mygrey}{gray}{0.35}
\definecolor{myblue}{rgb}{0.2,0.2,0.8}
\definecolor{myzard}{cmyk}{0,0,0.05,0}
\definecolor{mywhite}{rgb}{1,1,1}
\definecolor{mywhite}{rgb}{1,1,1}
\definecolor{myred}{rgb}{1,0.,0.3}

\def\be{ \begin{equation}}
\def\ee{ \end{equation}}
\def\bse{  \begin{subequations}}
\def\ese{  \end{subequations}}
\def\bea#1\ea{\begin{align}#1\end{align}}

\def\bi{\begin{itemize}}
\def\ei{\end{itemize}}

\def\bt{\begin{tabular}}
\def\et{\end{tabular}}

\def\ket#1{\vert #1 \rangle}

\def\half{\tfrac12}

\def\3half{\tfrac32}

\def\to{\rightarrow}

\def\U{\mathbf{U}}
\def\C{\mathbf{C}}
\def\H{\mathbf{H}}
\def\S{\mathbf{S}}
\def\C{\mathbf{C}}

\def\M{\mathbf{M}}

\def\Re{\text{Re}}
\def\Im{\text{Im}}
\def\R{\mathbf{R}}
\def\W{\mathbf{W}}
\def\M{\mathbf{M}}

\DeclareMathOperator*{\SumInt}{%
\mathchoice%
  {\ooalign{$\displaystyle\sum$\cr\hidewidth$\displaystyle\int$\hidewidth\cr}}
  {\ooalign{\raisebox{.14\height}{\scalebox{.7}{$\textstyle\sum$}}\cr\hidewidth$\textstyle\int$\hidewidth\cr}}
  {\ooalign{\raisebox{.2\height}{\scalebox{.6}{$\scriptstyle\sum$}}\cr$\scriptstyle\int$\cr}}
  {\ooalign{\raisebox{.2\height}{\scalebox{.6}{$\scriptstyle\sum$}}\cr$\scriptstyle\int$\cr}}
}

\begin{document}

\begin{abstract}
In this paper we propose a chiral resolution technique based on laser-induced continuum structure (LICS). We show that the two enantiomers can have a different ionization profile based on implemented LICS excitation. We treat a cyclic excitation between two bound states and a continuum state and show how asymmetric ionization can be achieved based on different trapping of the enantiomers due to the additional coupling between the bound states. Alternatively, when multiple states are involved in the interaction we investigate a multilevel LICS strategy of asymmetric ionization based on dark-bright state trapping. In this latter case one of the enantiomers is trapped in a dark state, which is immune to ionization, while the other is driven in a bright state, from which the population leaks into the continuum.
\end{abstract}

\author{K. N. Zlatanov}
\affiliation{Department of Physics, Sofia University, James Bourchier 5 blvd, 1164 Sofia,
Bulgaria}
\affiliation{Institute of Solid State Physics, Bulgarian Academy of Sciences, Tsarigradsko chaussée 72, 1784 Sofia, Bulgaria}
\author{N. V. Vitanov}
\affiliation{Department of Physics, Sofia University, James Bourchier 5 blvd, 1164 Sofia,
Bulgaria}
\title{Chiral resolution based on laser-induced continuum structure}
\date{\today }
\maketitle


\section{Introduction}

Chiral molecules are intriguing systems of great importance for both science and industry~\cite{Collins1993}. They play a vital role in biological processes, thus the  production of medical drugs often requires enantiopurity~\cite{Brooks2011}. This is due to the fact that usually only one of the enantiomers has beneficial health effect, while the other is inactive at best, if not harmful~\cite{Smith2009}. Due to the mirror-image structure of the enantiomers  their physical properties are very similar, which makes their detection and separation difficult. However their electric dipole interactions can differ and are consequently exploited by  optical based detection and separation techniques. For example, initially introduced by Kral and Shapiro~\cite{Kral2001} and experimentally demonstrated by Patterson~\cite{Patterson2013}, three-wave mixing has become one of the most prominent techniques that can selectively populate a specific state in only one of the enantiomers. 
This  selectivity serves as a solid foundation for plenty of detection and separation protocols~\cite{Li2010,Jacob2012,Shubert2015,Shubert2016,Leibscher2019,Wu2019,Vitanov2019,Chen2020,Xu2020}. Alternative chiroptical techniques are based on light induced forces~\cite{Donato2014,Bradshaw2015a,Bradshaw2015b,Ali2020}, selective dimer formation~\cite{Eilam2013} and dynamic Stark shift~\cite{Ye2019}. In addition optical techniques were proposed for  asymmetric synthesis~\cite{Zhdanov2007,Zhdanov2010} and chiral purification~\cite{Thomas2019,Gerbasi2006}. 
Ionization of the enantiomers is another feature that is known to generate asymmetric behaviour. For example, upon ionization with circularly polarized light the photoelectrons form the enantiomers experience different angular distribution, also known as photoelectron circular dichroism~(PECD), which is another widely used tool for chiral signal detection~\cite{Janssen2014,Artemyev2015,Muller2018,Muller2020,Ordonez2019,Beaulieu2018}. 

In this paper we investigate the possibility of enantio-separation based on asymmetric ionization. For that purpose we investigate the behaviour of chiral molecules under laser-induced continuum structure~(LICS). The LICS effect, reviewed in~\cite{Knight1990}, consist of a Raman-type linkage in which two or more bound levels are coupled to a common continuum state. It has been studied both theoretically~\cite{Armstrong1975,Coleman1982,Carroll1992,Carroll1993,Carroll1995,Carroll1996,Nakajima1994,Paspalakis1997,Paspalakis1998,Vitanov1997,Unanyan1998,
Rangelov2007,Nakajima2005,Stefanatos2021}, and experimentally, in atomic~\cite{Halfmann1998,Yatsenko1997,Yatsenko1999,Peters2005,Peters2007,Bohmer2002} and molecular systems~\cite{Faucher1999}.
The peculiarity of this type of system is that the total ionization probability experiences an extremum, in the same way as a Fano resonance~\cite{Fano1961}, when plotted as a function of the two-photon detuning. Whenever this extremum is a minimum the population is said to be "trapped", distributed predominantly among the bound states. The conditions under which trapping can occur depend both on the structural parameters of the system, and the control parameters of the field, such as frequency, intensity, time behaviour and chirping. By adding an additional coupling field in the LICS system, thus forming cyclic excitation between the two bound states and a continuum state, we show how the two enantiomers can experience different trapping conditions. Consequently predominant ionisation of one enantiomer over the other is expected to occur, thus allowing for charge based enantioseparation.  

Traditionally, the LICS effect has been applied in systems where two bound states can be isolated. In molecules such isolation might prove challenging due to the manifold of rotational and vibrational states that accompany the electronic structure. For such systems we propose an alternative multiple level LICS excitation which aims to induce asymmetric ionization based on dark-bright state trapping, that is driving one enantiomer in a dark state, which is immune to ionization, while setting the other in a bright state that is allowed to leak into the continuum. 
Here we investigate the exact condition under which such procedures can occur and outline under what initialization of the system the effect can be maximized. 

This paper is organized as follows.
In Section~\ref{sec:two} we introduce the LICS system and its properties.
Section ~\ref{sec:three} treats asymmetric ionization based on different trapping condition due to cyclic excitation of a LICS embedded system. Section ~\ref{sec:four} proposes an alternative LICS strategy for the case of multiple bound states based on dark-bright state trapping.
We conclude our presentation in Section~\ref{Sec:conclusion}.


\section{Dynamics of the LICS system}\label{sec:two}


\begin{figure}[tb]
\bt{c}
\includegraphics[width=0.55\columnwidth]{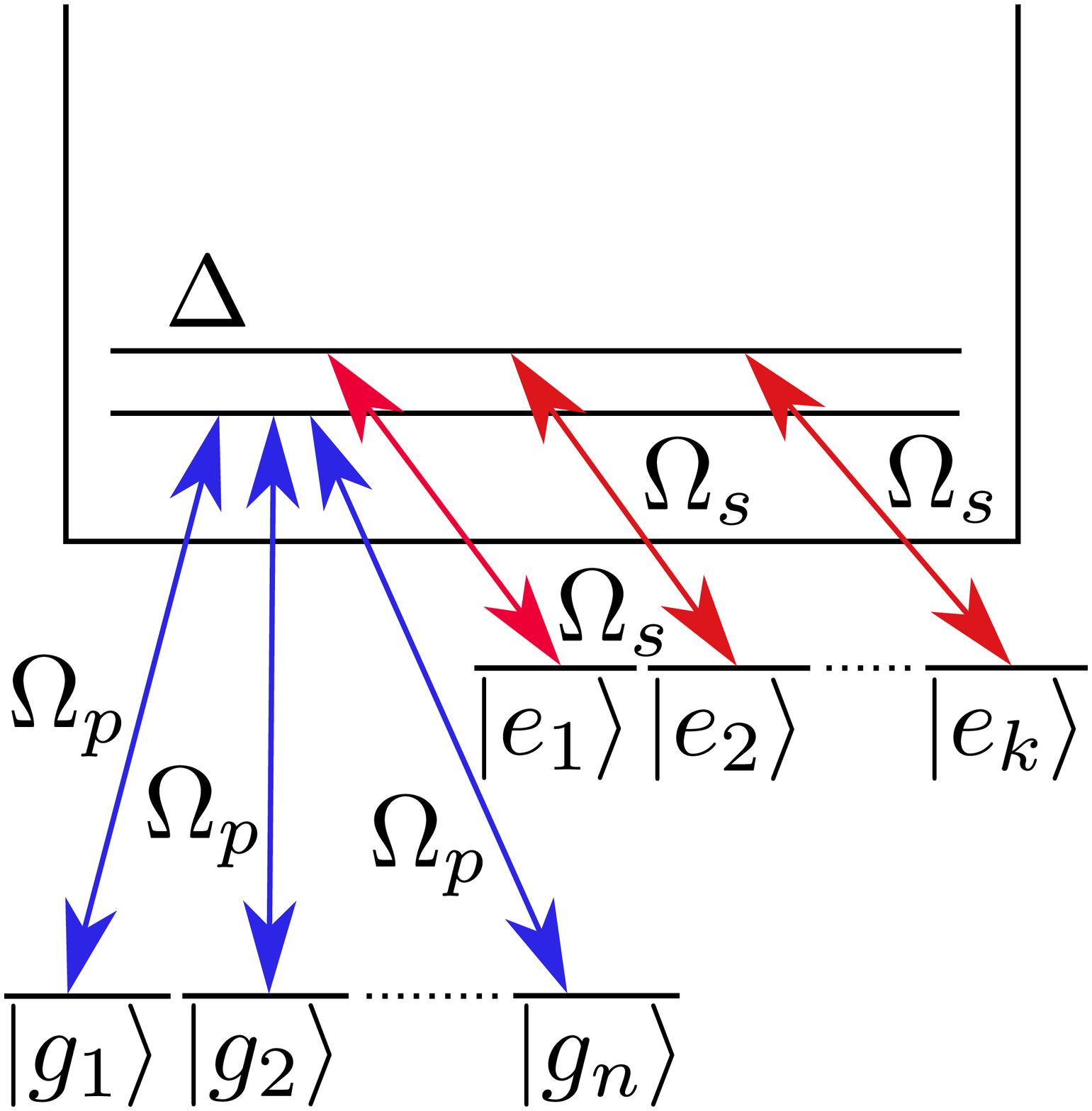}\llap{
  \parbox[b]{4.0 in}{(a)\\\rule{0ex}{1.61in}
  }}\\\\ \\ \includegraphics[width=0.55\columnwidth]{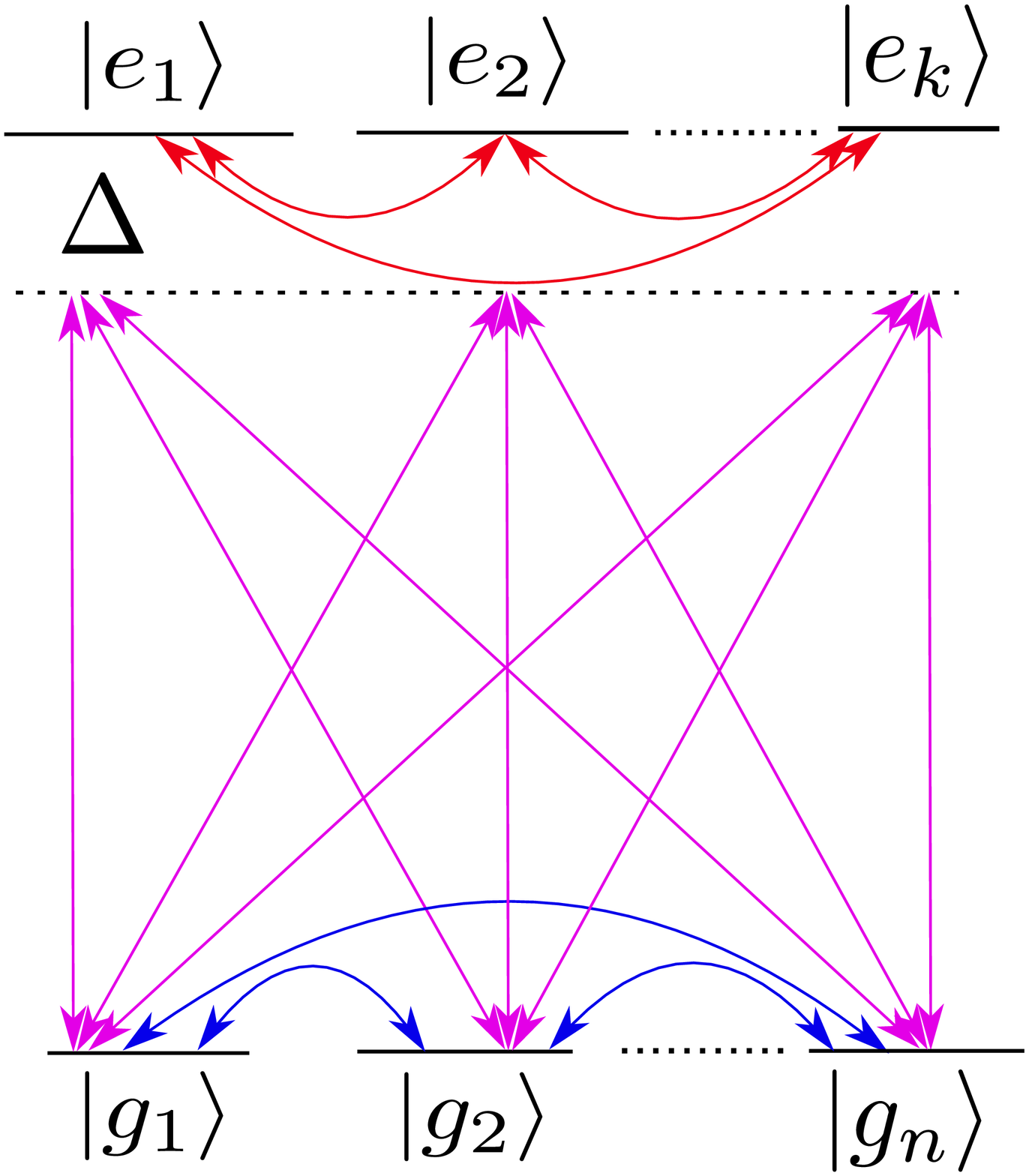}\llap{
  \parbox[b]{4.0 in}{(b)\\\rule{0ex}{1.61in}
  }}
\et
\caption{(Color online)
Multilevel LICS coupling scheme in (a) and its reduced system in (b) after elimination of the continuum (see text).
}\label{fig:System}
\end{figure}

A system composed of two bound levels with $n$ ground and $k$ excited degenerate states coupled through a common continuum in a Raman type linkage is described by the time dependent Schrodinger equation, which reads ($\hbar=1$)~\cite{Knight1990}, 
\bea \label{ContEqs}
i \frac{d}{dt}a_{g_i}(t)=\omega_{g_i} a_{g_i}(t)&-\int\displaylimits_{0}^{\infty} \Omega_{g_i\epsilon,p}(t) \cos(\omega_p t)a_{\epsilon}(t) \mathrm{d}\epsilon 
,\notag\\
\vdots\notag\\
i \frac{d}{dt}a_{e_j}(t)=\omega_{e_j} a_{e_j}(t)&-\int\displaylimits_{0}^{\infty} \Omega_{e_j\epsilon,s}(t) \cos(\omega_s t)a_{\epsilon}(t) \mathrm{d}\epsilon 
,\notag\\
\vdots\\
i \frac{d}{dt}a_{\epsilon}(t)=\omega_{\epsilon}a_{\epsilon}(t) &-\sum_{g_i}a_{g_i}(t) \Omega_{g_i\epsilon,p}(t) \cos(\omega_p t) 
\notag\\&-\sum_{e_j} a_{e_j}(t)  
\Omega_{e_j\epsilon,s}(t) \cos(\omega_s t),\notag
\ea
where $g_i$ runs over all the states in the bound ground level and $e_j$ runs over all the states in the bound excited level, and $\Omega_{n,\epsilon,l}$ are the Rabi frequencies connecting the $n$-th state with the continuum state $\ket{\epsilon}$ by the $l$-th laser, pump for ground-continuum transitions and Stokes for the excited-continuum transitions.  
The continuum state can be eliminated by formal integration of the last equation of Eqs.~(\ref{ContEqs}) and substituting it back in the equations for the amplitudes of the bound states. Thus with a change of the phase picture
 $a_n(t) \to c_n(t) =a_n(t)\exp(i \omega_n t)$ we can re-write the system of equations in matrix form,
\be
i\frac{d}{dt}\mathbf{C}(t)=\mathbf{H}(t)\mathbf{C}(t).  \label{TDSH}
\ee%
The Hamiltonian can be written in a block-matrix form as, 
 \be\label{H_general}
\H=-\half\left[
\begin{array}{cc}
 \mathbf{\Delta_g} & \mathbf{\Omega}\\
  \mathbf{\Omega}^{T} & \mathbf{\Delta_e}
\end{array}
\right]
\ee
with sub-matrices
\bse\label{H_elements}
\be
\Delta_{g_{ii}}=-2 S_g+i \Gamma_g,
\ee
\be
\Delta_{g_{ij}}=(q_{gg}+i)\Gamma_g, \quad i\neq j
\ee
\be
\Delta_{e_{ii}}=-2 S_e-2\Delta +i \Gamma_e, 
\ee
\be
\Delta_{e_{ij}}=(q_{ee}+i)\Gamma_e, \quad i\neq j
\ee
\be
\Omega_{ij}=(q_{ij}+i)\Gamma_{ij}.
\ee
\ese

The quantities in Eqs.~\eqref{H_elements} are defined as follows.
First of all, the Stark shifts caused by the lasers are defined as
\be
S_k= -\mathcal{P.V.} \displaystyle\SumInt d\epsilon\frac{\left|\Omega _{k,\epsilon,l}\right|^2}{4 (E_{\epsilon}-E_k-\omega_l)},
\ee
where $k$ runs over the bound states $g$ and $e$, $l$ runs over the lasers, $E_{\epsilon}$ designates the energy of the continuum state, $E_k$ is the energy of the respective bound state and $\omega_l$ is the frequency of the driving laser and $\mathcal{P.V.}$ stand for the principal value of the integral. 
The single-laser ionization rate is given by 
\be
\Gamma _{k} = \frac{1}{2} \pi    \left|\Omega _{k,\epsilon,l}\right|^2.\label{Gama_sl}
\ee
The two-photon coupling between the bound states due to the action of both lasers reads in a similar fashion
\be
\Gamma_{ij}=\frac{1}{2} \pi    \Omega _{i,\epsilon,a} \Omega _{j,\epsilon,b}=\sqrt{\Gamma_i\Gamma_j},\quad i\neq j.
\ee
Although eliminated from the equations, the continuum affects the evolution via the Fano parameters
\be
q_{ij}= \frac{\mathcal{P.V.} \displaystyle\SumInt d\epsilon\frac{\Omega _{i,\epsilon ,l} \Omega _{j,\epsilon,m}^*}{2 (E_{\epsilon}-E_g-\omega_l)}}{\Gamma_{ij}},
\ee
where $l$ and $m$ run over the lasers which drive the respective bound $\to$ continuum transition.
Due to the degeneracy of the system we distinguish between three different Fano parameters related to transitions via the continuum, namely (i) $q_{gg}$ for transitions linking $\ket{g_i}$ and
$\ket{g_j},$ (ii) $q_{ee}$ for transitions linking $\ket{e_m}$ and $\ket{e_n},$ and (iii) $q_{ge}$ transitions between $\ket{g_i}$ and $\ket{e_j}.$ 	
Finally, the two-photon detuning between the bound states reads
\be\label{detuning}
\Delta=E_{e_i}-E_{g_i}+\omega_{s}-\omega_{p}.
\ee
The reduced system is illustrated in Fig.~\ref{fig:System} (b). 

With the intent to show how LICS can be used for enantioseparation we proceed forward to two different strategies that illustrate our concept in the next sections. 
\section{Enantioselective  ionization among two bound states}\label{sec:three}
\begin{figure}[tb]
\bt{c}
\includegraphics[width=0.60\columnwidth]{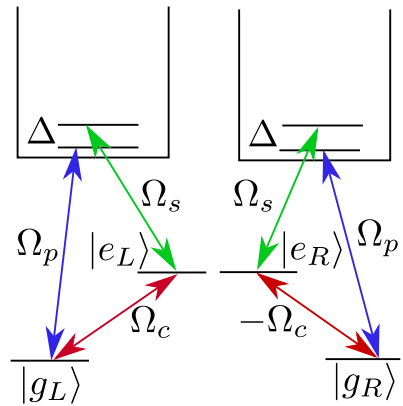}
\et
\caption{(Color online)
Cyclic LICS excitation for the $L$ enantiomer (left frame) and the $R$ counterpart (right frame).}
\label{fig:2lvl system}
\end{figure}

We begin by considering only two bound states $\ket{g}$ and $\ket{e}$~\cite{Knight1990} coupled to a common continuum with continuous wave (cw) lasers. The Hamiltonian of Eq.~(\ref{H_general}) takes the form
\be\label{H2lvl}
\H=\half\left[
\begin{array}{cc}
 2 S_g-i \Gamma_g & -(q+i)\Gamma_{ge}\\
  -(q+i)\Gamma_{ge} & 2 S_e+2\Delta -i \Gamma_e
\end{array}
\right].
\ee
The eigenvalues of this Hamiltonian are complex and for that reason upon excitation some of the population will leak through the continuum and some will be "trapped" in the bound states. In order to maximize the bound state population we have to drive the system in a way that will prevent this leakage, by imposing conditions on the control parameters. Mathematically this requires at least one eigenvalue to be real, which is ensured if we solve the eigenvalue equation 
\be\label{lambdaIM}
\Im \Big[\det\left(\H-\lambda\mathbb{1}\right)\Big]=0,
\ee
for $\lambda$ and then impose conditions on the detuning such that 
\be
\Re \Big[\det\left(\H-\lambda\mathbb{1}\right)\Big]=0,
\ee
holds, once we substitute $\lambda$ from Eq.~(\ref{lambdaIM}). For two bound states this yields~\cite{Paspalakis1997}
\be\label{2lvl trapping}
\Delta=\frac{1}{2} q \left(\Gamma _g-\Gamma _e\right)+S_g-S_e.
\ee
All of the quantities in Eq.~(\ref{2lvl trapping}) are controlled by the laser pulses, e.g. the frequencies determine the detuning, while the right-hand side is controlled by the intensity.  
Naturally the ionization of the system 
\be
I=1-\sum_i |c_i|^2,
\ee
will experience a minimum when Eq.~(\ref{2lvl trapping}) is satisfied.  This type of excitation can be exploited 
for asymmetric ionization of chiral molecules if we ensure that the two enantiomers experience different trapping condition. This can be achieved by employing an additional coupling $\Omega_c$ between the bound states as shown in Fig.~\ref{fig:2lvl system}, which is of opposite sign for the enantiomers, when all three fields have different, mutually perpendicular polarization. 

The transformed Hamiltonian of Eq.~(\ref{H2lvl}) will now read
\be\label{H_cyclic}
\H_{cyclic}=\half\left[
\begin{array}{cc}
  2 S_g-i \Gamma _g &  \pm\Omega _c-(q+i) \Gamma
   _{\text{ge}}\\
 \pm\Omega _c-(q+i) \Gamma _{\text{ge}} & 2 \Delta
   +2 S_e-i \Gamma _e\\
\end{array}
\right],
\ee
with $\pm$ standing for the $L$ and $R$ molecules respectively.
 
With cyclic three-wave coupling of the system, we ensure that the two enantiomers experience different trapping condition, namely,
\be\label{cyclic trapping}
\Delta=\delta S -\frac{\left(\Gamma _e-\Gamma _g\right) \left(q \Gamma _e \Gamma _g\pm\Omega _c \Gamma
   _{\text{ge}}\right)}{2 \Gamma _e \Gamma _g},
\ee
where $\delta S=S_g-S_e$ is the difference of the Stark shifts of the bound states and the $\pm$ stands for the $L$ and $R$ enantiomers respectively.
The control laser generates its own Stark shift, namely
\be
S_c=-\frac{1}{4}\sum_m\frac{\Omega_{nm}\Omega_{mk}}{E_m-E_n-\omega_c}+\mathcal{O}(\Omega_p \Omega_s \Omega_c),
\ee
with $m,n$ and $k$ running over all the states of the molecule. As a consequence the $\delta S$ term can also be different for the two enantiomers. In general the Stark shift of each state
\bea
S_j&= S_{j,p}+S_{j,s}+S_{j,c}, \quad j=g,e
\ea 
contains terms $\mathcal{O}(\Omega_p \Omega_s \Omega_c),$ which will differ between the two enantiomers and the overall $\delta S$ term will be proportional to the control field~\cite{Ye2019,Lehmann2015}. 
Thus the asymmetry of the trapping is controlled by the strength of the $\Omega_c$ field as it appears explicitly in Eq.~(\ref{cyclic trapping}) and also implicitly affects the Stark shifts. 
\begin{figure}[tb]
\bt{c}
 \includegraphics[width=0.85\columnwidth]{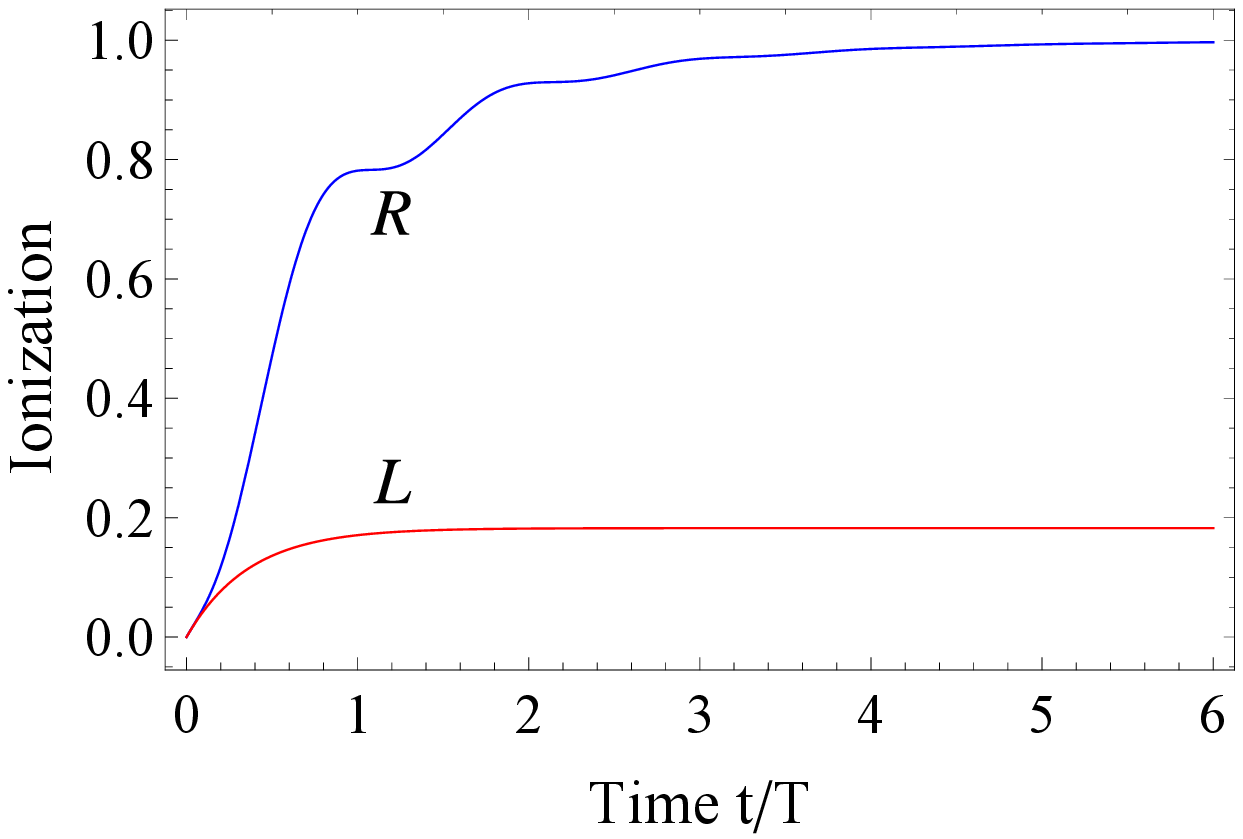}\llap{
  \parbox[b]{4.3 in}{(a)\\\rule{0ex}{1.71in}}}
\vspace{5mm}   
   \\ \includegraphics[width=0.85\columnwidth]{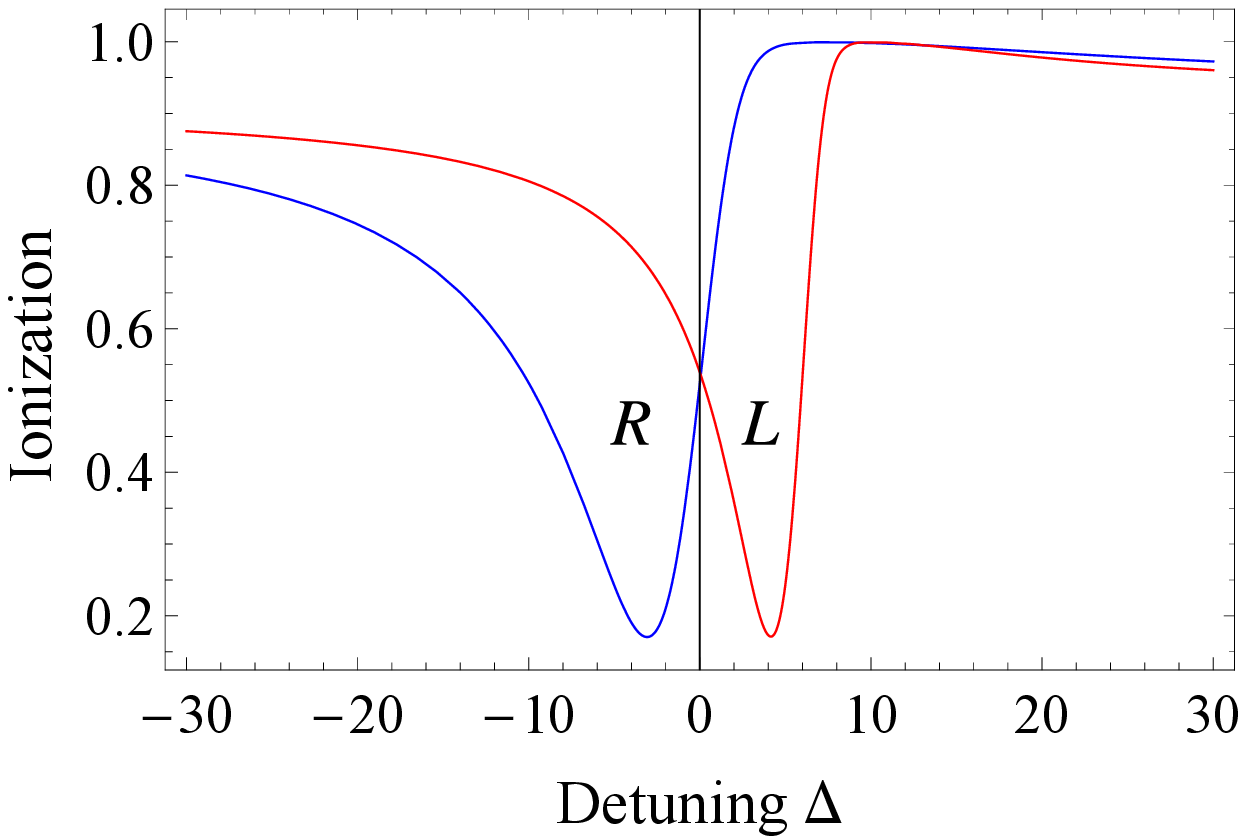}\llap{
  \parbox[b]{4.3 in}{(b)\\\rule{0ex}{1.71in}
  }}
\et
\caption{(Color online)
Ionization probability for the two enantiomers with respect to normalized time $t/T$ (a) and their Fano profiles (b) as a function of the two-photon detuning $\Delta$~(see Eq.~(\ref{detuning})). 
Both frames have the same excitation parameters set to $\Gamma_g=0.5 T^{-1}, \Gamma_e=2.24T^{-1}, \Omega_c=1.2 T^{-1}, q=4, \delta S_{L}=7 T^{-1}, \delta S_R=2 T^{-1}.$ In (a) the detuning is $\Delta=4.506T^{-1},$ corresponding to the minimum of the $L$ enantiomer in (b) while the Fano profiles of (b) are at $t/T=5.$
}\label{fig:2lvl profiles}
\end{figure}
This system has an analytical solution which is too cumbersome to be shown here. Instead we present it graphically (for system initialized in the ground bound state) in
Fig.~\ref{fig:2lvl profiles}(a), which illustrates the ionization of the enantiomers in time normalized to $T,$ which can be any appropriate time scale, e.g. if pulsed excitation is invoked (we treat cw excitation) it will be the pulse duration, alternatively it can be the ionization saturation time, or the time for complete ionization. 
Fixing the detuning such that Eq.~(\ref{cyclic trapping}) is satisfied only for the $L$ enantiomer, the $R$ counterpart experiences a predominant ionization that grows in time. The way we choose the detuning in order to trap the $L$ or the $R$ enantiomer is not symmetric. This can be seen from the Fano profile given in Fig.~\ref{fig:2lvl profiles}(b). The two profiles differ due to the $\pm \Omega_c$ field.  When we fulfil the trapping condition such that the $L$ enantiomer is trapped we are under the "high wing" of the Fano profile of the $R$ enantiomer. If instead we choose to fulfil the trapping condition for the $R$ enantiomer then the $L$ counterpart will have a lower ionization probability. It is worth mentioning that the asymmetric Fano profile is characteristic for specific intensity regime, e.g. if one of the lasers is significantly  stronger than the other this may generate an asymmetric Fano profile, which is typical in the short interaction times. If we look at longer interaction times, the profiles will become more symmetric, and consequently it does not make much of a difference which enantiomer will be trapped. 

\section{Enantioselective ionization among multiple bound states}\label{sec:four}
The cyclic LICS model of the previous section can separate chiral molecules whenever only two bound states can be isolated. Experimentally it is often challenging to achieve such isolation, especially in molecular systems where rotational and vibration states are present. As we have shown in a previous paper~\cite{Zlatanov2021}, multiple bound states distort the LICS evolution drastically. Therefore in this section we address this type of problem.
Unlike the previous approach where we employed an additional coupling between the bound states since we needed the cyclic excitation, here we will consider a Raman type transition between five ground and five excited bound states coupled to the continuum by cw lasers. This system is justified if we take, for example, the well studied Rydberg states of fenchone~\cite{Goetz2017} as a possible realization of our proposed approach.
The Hamiltonian of the system is given by Eq.~(\ref{H_general}) and we assume it to be degenerate. Still after the elimination of the continuum it is a $10\times10$  matrix out of which we can not extract analytically much information about the evolution of the system. For that matter it is reasonable to change basis in a way that a block-diagonal structure can appear, thus simplifying the problem to independent blocks of smaller dimension. Such transformation can be carried out by a sequence of Givens rotations, which rotate the basis between any two states, for example 
\be
\R(\alpha)=\left[
\begin{array}{cc}
 \cos (\alpha) & \sin (\alpha ) \\
 -\sin (\alpha ) & \cos (\alpha ) \\
\end{array}
\right],
\ee
will rotate a state and its nearest neighbour, while
\be
\widetilde{\R}(\beta)=\left[
\begin{array}{ccc}
 \cos (\beta ) & 0 & \sin (\beta ) \\
 0 & 1 & 0 \\
 -\sin (\beta ) & 0 & \cos (\beta ) \\
\end{array}
\right]
\ee
will rotate a state and its second nearest neighbour. 
We can construct a unitary transformation matrix by two or more Givens rotations acting on the ground and excited states simultaneously, for example
\be\label{U alpha1}
\U(\alpha)=\left[
\begin{array}{cccc}
 \R(\alpha)_{2\times2} &\hdots &\hdots & 0 \\
  \vdots & \mathbb{1}_{3\times3} & 0 & \vdots \\
     	 \vdots & 0 & \R(\alpha)_{2\times2} &\vdots  \\
     	 0 & \hdots &\hdots &\mathbb{1}_{3\times3}
\end{array}
\right]_{10 \times 10},
\ee
thus a composite rotation can be employed for the basis change that will get the Hamiltonian in block diagonal form.
In our current $10\times10$ example this will read
\be
\W(\alpha_1,\alpha_2,\beta,\alpha_3,\alpha_4)=\S\U(\alpha_4)\U(\alpha_3)\U(\beta)\U(\alpha_2)\U(\alpha_1),
\ee
where the $\U(\alpha_1)$ has the structure of Eq.~(\ref{U alpha1}), while the other four transformations are given as,
\bse
\be
\U(\alpha_2)=\left[
\begin{array}{ccccc}
1 & 0 & \hdots &\hdots & 0 \\
0 & \R(\alpha_2)_{2\times2} & \hdots & \hdots & \vdots \\
\vdots  & \hdots & \mathbb{1}_{3\times3} &\hdots & \vdots \\
\vdots     	 &\hdots & \hdots & \R(\alpha_2)_{2\times2} & \vdots  \\
0 & \hdots & \hdots &\hdots & \mathbb{1}_{2\times2}
\end{array}
\right],
\ee 
\be
\U(\beta)=\left[
\begin{array}{cccc}
\widetilde{\R}(\beta)_{3\times3} & 0 & \hdots & 0 \\
0 & \mathbb{1}_{2\times2}  &\hdots & \vdots \\
\vdots  & \hdots & \widetilde{\R}(\beta)_{3\times3} & \vdots \\
0  &\hdots & \hdots & \mathbb{1}_{2\times2}
\end{array}
\right],\ee \\
\be
\U(\alpha_3)=\left[
\begin{array}{ccccc}
\mathbb{1}_{2\times2} & 0 & \hdots &\hdots & 0 \\
0 & \R(\alpha_3)_{2\times2} & \hdots & \hdots & \vdots \\
\vdots  & \hdots & \mathbb{1}_{3\times3} &\hdots & \vdots \\
\vdots     	 &\hdots & \hdots & \R(\alpha_3)_{2\times2} & \vdots  \\
0 & \hdots & \hdots &\hdots & 1
\end{array}
\right],\ee\\
\be
\U(\alpha_4)=\left[
\begin{array}{cccc}
\mathbb{1}_{3\times3} & 0 & \hdots & 0 \\
0 & \R(\alpha_4)_{2\times2}  &\hdots & \vdots \\
\vdots  & \hdots & \mathbb{1}_{3\times3} & \vdots \\
0  &\hdots & \hdots & \R(\alpha_4)_{2\times2}
\end{array}
\right],
\ee
\ese
and $\S$ is a shifting operator 
\be
\S=\left[
\begin{array}{cccccccccc}
 1 & 0 & 0 & 0 & 0 & 0 & 0 & 0 & 0 & 0 \\
 0 & 1 & 0 & 0 & 0 & 0 & 0 & 0 & 0 & 0 \\
 0 & 0 & 1 & 0 & 0 & 0 & 0 & 0 & 0 & 0 \\
 0 & 0 & 0 & 0 & 1 & 0 & 0 & 0 & 0 & 0 \\
 0 & 0 & 0 & 0 & 0 & 0 & 0 & 1 & 0 & 0 \\
 0 & 0 & 0 & 0 & 0 & 1 & 0 & 0 & 0 & 0 \\
 0 & 0 & 0 & 0 & 0 & 0 & 1 & 0 & 0 & 0 \\
 0 & 0 & 0 & 0 & 0 & 0 & 0 & 0 & 0 & 1 \\
 0 & 0 & 0 & 1 & 0 & 0 & 0 & 0 & 0 & 0 \\
 0 & 0 & 0 & 0 & 0 & 0 & 0 & 0 & 1 & 0 \\
\end{array}
\right],
\ee

that rearranges rows and columns. The rotation angles are given as follows
\bea
\alpha _1&= \frac{\pi }{2},\quad\alpha _2= \frac{\pi }{4},\quad\beta = -\tan ^{-1}
\left(\frac{1}{\sqrt{2}}\right),\notag\\\alpha _3&= -\frac{\pi
   }{3},\quad\alpha _4= \frac{1}{2} \tan ^{-1}\left(\frac{4}{3}\right).
\ea
Since we only use rotation matrices the unitarity relation $\W\W^{\dagger}=\mathbb{1}$ is automatically satisfied.
Then upon transformation of the Hamiltonian of Eq.~(\ref{H_general}) we achieve the desired block diagonal form
\be \label{MLICS WHW}
\W\H\W^{\dagger}=\left[
\begin{array}{cc}
 \H_d & 0 \\
 0 & \H_b \\
\end{array}
\right],
\ee
where $\H_b$ is a $2\times2$ bright Hamiltonian linking a set of ground bound states to a set of excited bound states
\bse
\be
\H_b=\half\left[
\begin{array}{cc}
 S_g-\left(4 q_{\text{gg}}+5 i\right) \Gamma _g & -5 \left(q_{\text{ge}}+i\right)
   \sqrt{\Gamma _e \Gamma _g} \\
 -5 \left(q_{\text{ge}}+i\right) \sqrt{\Gamma _e \Gamma _g} &  \Delta +S_e -\left(4 q_{\text{ee}}+5 i\right)
   \Gamma _e \\
\end{array}
\right],
\ee
 while $\H_d$ is a $8\times8$ dark Hamiltonian 
\be
\H_d=\left[
\begin{array}{cccccc}
\varepsilon_g & 0 & \hdots & \hdots & \hdots & 0\\
0 & \ddots & 0 & \hdots & \hdots & 0 \\
\vdots & 0 & \varepsilon_g & \hdots & \hdots & 0 \\
\vdots & \hdots &  \hdots  &  \varepsilon_e & \hdots & 0\\
\vdots & \hdots &  \hdots  &  \hdots & \ddots & 0\\
0  & \hdots  &  \hdots & \hdots & \hdots & \varepsilon_e 
\end{array}
\right]_{8\times 8},
\ee
\ese
driving the evolution of four ground and four excited dark states with energies 
\bea
\varepsilon_g & = \frac{1}{2} \left(2S_g+\Gamma _g q_{\text{gg}}\right),\\
\varepsilon_e &= \frac{1}{2} \left(2 \Delta +2S_e+\Gamma _e q_{\text{ee}}\right).
\ea
The states in the rotated basis are now superposition sets of the ground and excited bound states,
\be \label{MLICS vec}
\W \C(t)=\left[
\begin{array}{c}
-\frac{1}{\sqrt{6}}\left[c_{g_1}-2 c_{g_2}+c_{g_3}\right]\\ 
 \frac{1}{\sqrt{2}}\left[c_{g_3}-c_{g_1} \right] \\
   \frac{1}{2 \sqrt{3}}\left[c_{g_1}+c_{g_2}+c_{g_3}-3 c_{g_4} \right] \\
   -\frac{1}{2 \sqrt{5}}\left[c_{g_1}+c_{g_2}+c_{g_3}+c_{g_4}-4 c_{g_5} \right]\\
   \frac{1}{2 \sqrt{3}}\left[c_{e_1}+c_{e_2}+c_{e_3}-3 c_{e_4} \right] \\ 
   -\frac{1}{\sqrt{6}}\left[c_{e_1}-2 c_{e_2}+c_{e_3} \right]\\ 
    \frac{1}{\sqrt{2}}\left[c_{e_3}-c_{e_1} \right]\\
   -\frac{1}{2 \sqrt{5}} \left[c_{e_1}+c_{e_2}+c_{e_3}+c_{e_4}-4 c_{e_5} \right]\\
   \frac{1}{\sqrt{5}} \left[c_{g_1}+c_{g_2}+c_{g_3}+c_{g_4}+c_{g_5} \right] \\
   \frac{1}{\sqrt{5}}\left[c_{e_1}+c_{e_2}+c_{e_3}+c_{e_4}+c_{e_5}\right] 
\end{array}
\right].
\ee
A closer inspection of Eq.~(\ref{MLICS vec}) reveals that the number of states in the superpositions range from $2$ to the number of bound states in the ground or excited levels; in our case both are five. In fact this result can be generalized for this type of system of any dimension. The technical reason lies within the structure of the composite rotation, which is of block diagonal form (before the shifting operator) and the blocks resemble triangular structure.
 
The way we can use this system for asymmetric ionization is by initially preparing, for example, the $R$ enantiomer, say with the cyclic excitation technique of Appendix ~\ref{app:Cyclic superposition}, in the dark state of smallest dimension,
\bse\label{MLICS superpositon}
\be\label{MLICS R-state}
\ket{c_R}=\frac{1}{\sqrt{2}}\left[\ket{c_{g_3}}-\ket{c_{g_1}} \right],
\ee
while the other will be prepared in
\be \label{MLICS L-state}
\ket{c_L}=\frac{1}{\sqrt{2}}\left[\ket{c_{g_3}}+\ket{c_{g_1}} \right].
\ee
\ese
\begin{figure}[tb]
\bt{c}
\includegraphics[width=0.85\columnwidth]{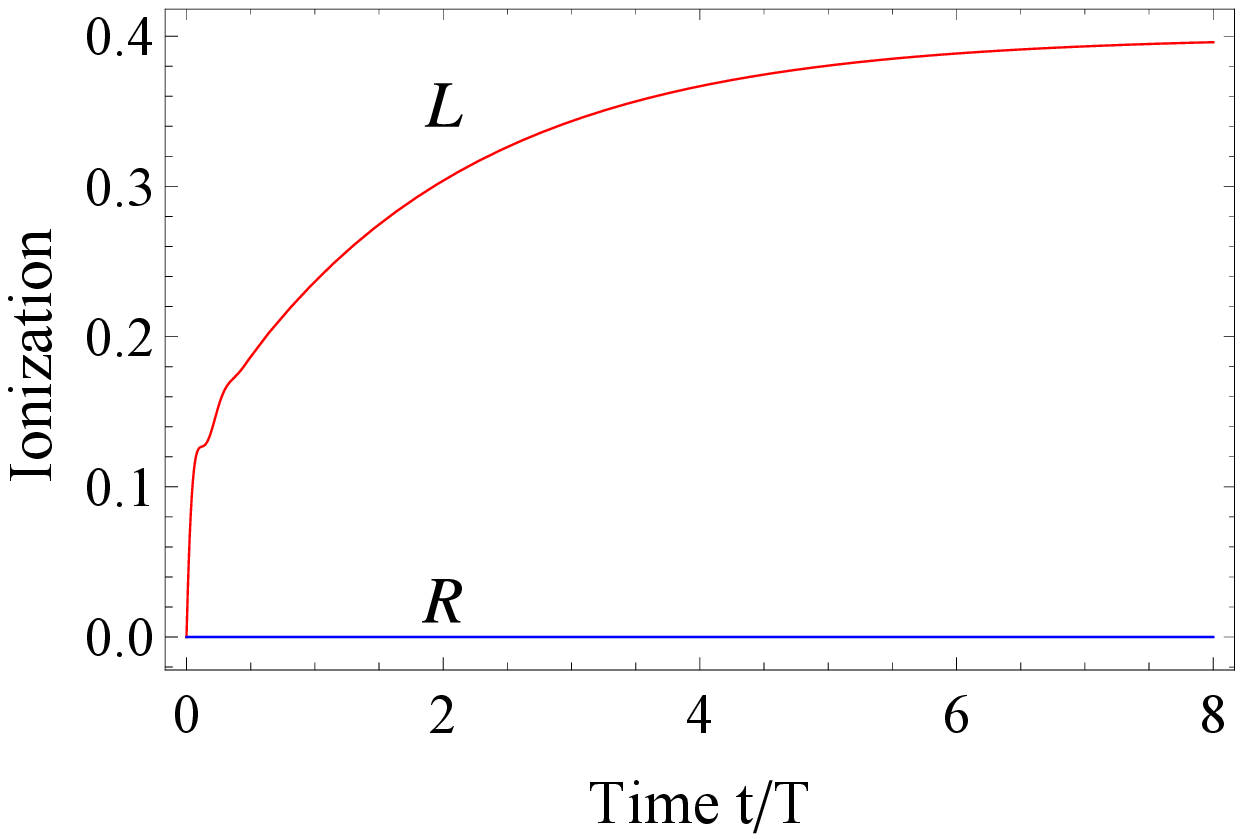}\llap{
  \parbox[b]{4.25 in}{(a)\\\rule{0ex}{1.651in}
  }} \vspace*{5mm} \\ \includegraphics[width=0.85\columnwidth]{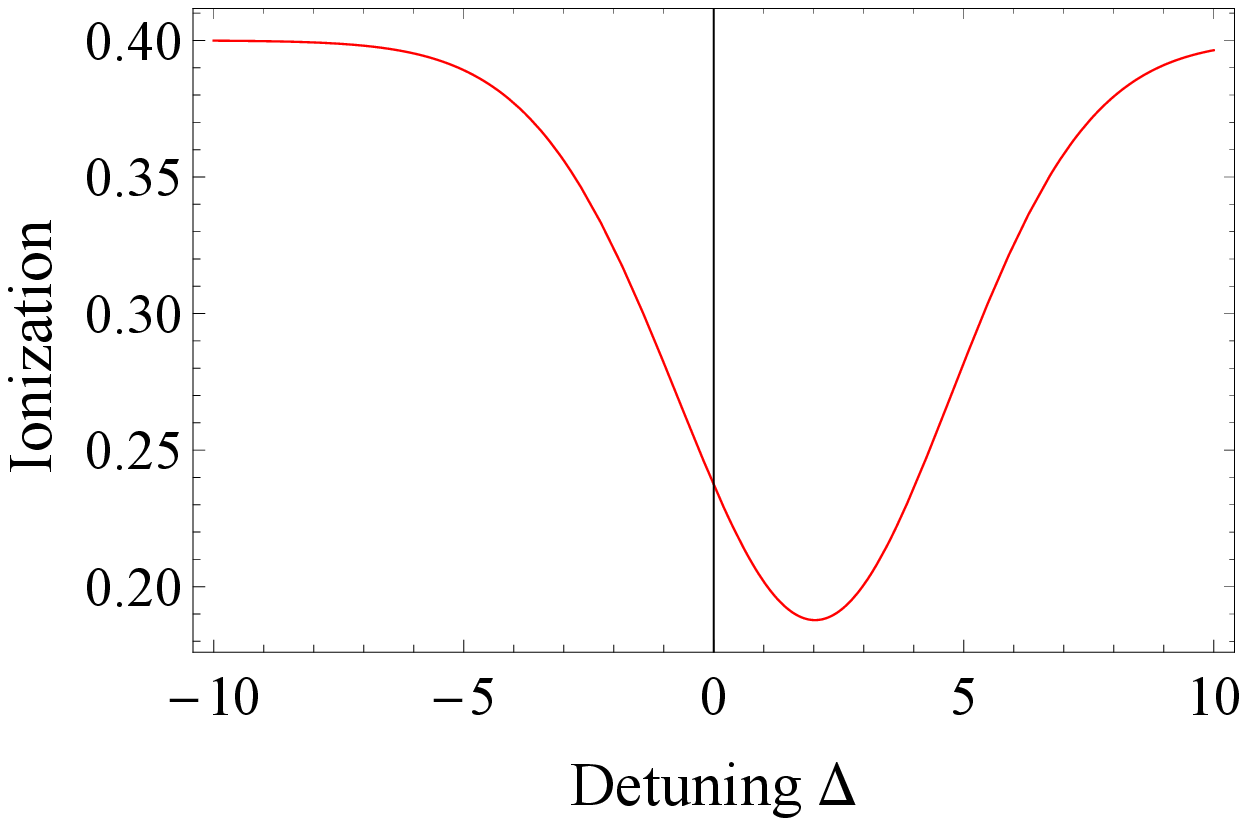}\llap{
  \parbox[b]{4.25 in}{(b)\\\rule{0ex}{1.601in}}}
\et
\caption{(Color online)
Same as in Fig.~\ref{fig:2lvl profiles} but for the multilevel LICS system of Fig.~\ref{fig:System}.  In (a) the $R$ enantiomer is in the dark state of Eq.~(\ref{MLICS R-state}), while the initialization of the $L$ counterpart in the state of Eq.~(\ref{MLICS L-state}) populates the bright state and additional three dark states. In order to maximize the ionization of the $L$ enantiomer we want to be away from the Fano minimum of the (b) frame. 
Both frames have the same excitation parameters set to $\Gamma_g=1.7 T^{-1}, \Gamma_e=1.9T^{-1}, q_{gg}=1.2, q_{ee}=2.4, q_{ge}=2.26, S_{g}=19 T^{-1}, S_e= 20T^{-1}$. In (a) the detuning is $\Delta=-6.2T^{-1}$ and the Fano profiles of (b) are at $t/T=8.$
}\label{fig:MLICS profiles}
\end{figure}
The idea here relies on the fact that the dark-state energy is real; therefore the dark state will not get ionized due to the coherent interference.
Since the state of Eq.~(\ref{MLICS L-state}) is not an eigenstate of the rotated basis the evolution of the $L$ enantiomer will be driven by the full Hamiltonian of Eq.~(\ref{MLICS WHW}). This will translate into a growing ionization since the population will leak through the bright states into the continuum as shown in Fig.~\ref{fig:MLICS profiles} (a).
Note that the ionization saturates as it reaches $40\%,$ as we can see from the figure. This is not a consequence of the specific excitation parameters but rather streams from the system itself. The reason is that as we initialize the $L$ molecule in the state of Eq.~(\ref{MLICS L-state}) we populate the bright ground state, as well as some of the dark states. The latter preserve the population from being ionized through the coherent channels. For our purposes, which are to maximize the ionization of the $L$ enantiomer, we need to set the detuning away from the Fano minimum shown in Fig.~\ref{fig:MLICS profiles}(b). It is important to note that initializing the system in any other dark state besides the state with smallest amount of components will result in additional lowering of the ionization probability, that is away from the Fano minimum, due to the larger number of dark states that will participate in the interaction.

\section{Discussion and Summary}\label{Sec:conclusion}
In this paper we explored enantioselective ionization of chiral molecules based on laser-induced continuum structure, employing two different strategies. When we embed LICS in a cyclic excitation the ionization asymmetry derives from the different trapping conditions experienced by the enantiomers. Setting the detuning in a way that one of the enantiomers is trapped, its ionization is in a minimum, while the others can be strongly elevated or in a maximum. The distance between the Fano minima, and consequently between a minimum and a peak, of the two enantiomers, is explicitly and implicitly (due to the Stark shifts) controlled by the control field $\Omega_c.$  Whenever two bound states can not be isolated from the rest of the Hilbert space, we employed a different strategy that relies on dark-bright state trapping, that is setting one enantiomer in a dark state, thus blocking the ionization, while the other in a bright state which leaks into the continuum. In our example the ionization of the enantiomer in a bright state reached $40\%,$ while the enantiomer in the dark state experienced no ionization. 

We note that in our idealized model we do not account for incoherent channels for ionization. These channels are always present, so naturally we expect that even the enantiomers trapped in the dark states will experience some ionization, but it will be rather small as it can be suppressed both by picking the proper parameters~\cite{Stefanatos2021} or by other techniques~\cite{Unanyan1998}.

Our proposed implementation of LICS resembles resonance-enhanced multiphoton ionization~(REMPI) in a way, and in a certain scenario REMPI can induce a LICS effect. For example in cases where two-colour (2+1) REMPI is employed, where the pumping and ionizing field act simultaneously, the overlap between the two fields can link the transition state through the continuum to a Rydberg state by another absorption from the pumping field thus generating LICS. Such is also the case if the transition state has multiple states nearby, which can all be coupled to the continuum by a Raman transition. In such cases our study can help for the optimization of ion yield. 
\appendix

\section{Generation of superposition states in cyclic system}\label{app:Cyclic superposition}
\begin{figure}[tb]
\includegraphics[width=0.5\columnwidth]{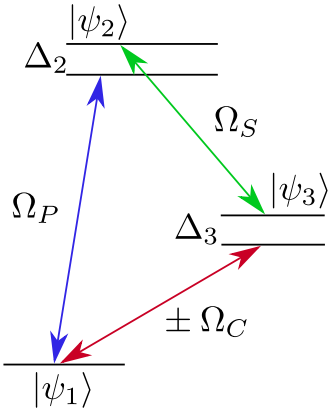}
\caption{(Color online)
Standard cyclic excitation between three bound states}\label{fig:appendix}
\end{figure}

We illustrate here how a the two enantiomers can be initialized in the states of Eqs.~(\ref{MLICS superpositon}). We consider a simplified version of the procedure, suited for our problem, for a detailed investigation of the system we refer the reader to~\cite{Rangelov2008}. The Hamiltonian of a three state system under cyclic excitation given in Fig.~\ref{fig:appendix} reads,
\be
\H_{3wave}=\half\left[
\begin{array}{ccc}
 0 & \Omega _P(t)e^{i\phi_p} &  \Omega _C(t) \\
  \Omega _P(t)e^{-i\phi_p} & 2 \Delta_1 & \Omega _S(t)e^{i\phi_s} \\
  \Omega _C(t) &  \Omega _S e^{-i\phi_s} & 2 \Delta_2 \\
\end{array}
\right].
\ee 

Upon transformation by 
\be
\M=\left[
\begin{array}{ccc}
 1 & 0 & 0 \\
  0 & \cos \theta &e^{-i\phi_p} \sin \theta \\
  0 & e^{i\phi_p} \sin\theta & -\cos \theta \\
\end{array}
\right],
\ee
where $\tan \theta=\Omega_C/\Omega_P,$ the new Hamiltonian 
\be
\widetilde{\H}=\M\H_{3wave}\M^{\dagger}-\rm{i}\M\dot{\M},
\ee now reads,  
\be\label{app: W hamiltonian}
\widetilde{\H}=\left[
\begin{array}{ccc}
 0 & \widetilde{\Omega }_P & 0 \\
 \widetilde{\Omega }_P^{*} & 2 \widetilde{\Delta }_2 & \widetilde{\Omega }_S-2 i e^{-i\phi_p} \dot{\theta } \\
 0 & \widetilde{\Omega }_S^{*}+2 i e^{i\phi_p} \dot{\theta } & 2 \widetilde{\Delta }_3 \\
\end{array}
\right].
\ee
The effective Rabi frequencies are given by

\bse
\bea
\widetilde{\Omega }_P=&e^{i\phi_p}\sqrt{\Omega_P^2+\Omega_S^2}=e^{i\phi_p}\Omega,\\
\widetilde{\Omega }_S=&\frac{1}{\Omega^2}\left[2e^{-i\phi_p}(\Delta_2-\Delta_3)\Omega_P\Omega_C\right.\notag\\ &\left.+\left(e^{-2i(\phi_p+\phi_s)}\Omega_C^2-\Omega_P^2 \right)e^{i\phi_s}\Omega_S\right],
\ea
\ese
and the detunings read
\bse
\bea
\widetilde{\Delta}_2&=\frac{\Delta_3\Omega_C^2+\Delta_2\Omega_P^2+\Omega_P\Omega_S\Omega_C\cos(\phi_p+\phi_s)}{\Omega^2},\\
\widetilde{\Delta}_3&=\frac{\Delta_2\Omega_C^2+\Delta_3\Omega_P^2+\Omega_P\Omega_S\Omega_C\cos(\phi_p+\phi_s)}{\Omega^2}.
\ea
\ese
The transformed states on the other hand  read,
\bse
\bea
\widetilde{\psi}_1&=\psi_1,\\
\widetilde{\psi}_2&=\psi_2 \cos \theta + \psi_3 \sin\theta e^{-i\phi_p},\\
\widetilde{\psi}_3&=\psi_2\sin\theta e^{i\phi_p} -\psi_3 \cos\theta. \label{app:psi 3}
\ea
\ese
When we pick the control parameters such that $\widetilde{\Delta}_3=0$ the transformed Hamiltonian of Eq.~(\ref{app: W hamiltonian}) resembles a lambda system, over which we can do STIRAP. Fixing the transformed Rabi frequencies in a counter-intuitive sequence, that is $\widetilde{\Omega }_S-2 i e^{-i\phi_p} \dot{\theta }$ before $\widetilde{\Omega}_P,$ a complete population transfer can occur between $\widetilde{\psi}_1$ and $\widetilde{\psi}_3.$ In the original basis this will constitute a superposition of Eq.~(\ref{app:psi 3}), thus setting $\pm\Omega_C/\Omega_P=\pm\pi/4$ we can initialize the two enantiomers in the states of Eq.~(\ref{MLICS superpositon}).

\acknowledgments
KNZ acknowledges support from the project MSPLICS - P. Beron Grant from The Bulgarian National Science Fund (BNSF).


\end{document}